\documentclass[%
 reprint,
 amsmath,amssymb,
 aps,
]{revtex4-2}
\usepackage{soul,xcolor}\usepackage{siunitx}
\setstcolor{red}
\usepackage{braket}
\usepackage{hyperref}
\hypersetup{colorlinks=true, allcolors=blue}
\usepackage{float}
\usepackage[super]{nth}
\usepackage{mathtools}
\usepackage{amsmath}

\usepackage{graphicx}
\usepackage{dcolumn}
\usepackage{bm}

\usepackage{soul}

\begin{document}

\preprint{APS/123-QED}

\title{Wigner Function Tomography via Optical Parametric Amplification}

\author{Mahmoud Kalash$^{1,2}$}
 \email{mahmoud.kalash@mpl.mpg.de}
\author{Maria V. Chekhova$^{1,2}$}%
\affiliation{$^{1}$Max Planck Institute for the Science of Light, Staudtstr. 2, 91058 Erlangen, Germany. \\
 $^{2}$Friedrich-Alexander Universit\"at Erlangen-N\"urnberg, Staudtstr. 7/B2, 91058 Erlangen, Germany.
}%
\date{\today}
\begin{abstract}
Wigner function tomography is indispensable for characterizing quantum states, but its commonly used version, balanced homodyne detection, suffers from several weaknesses. First, it requires efficient detection, which is critical for measuring fragile non-Gaussian states, especially bright ones. Second, it needs a local oscillator, tailored to match the spatiotemporal properties of the state under test, and fails for multimode and broadband states. Here we propose Wigner function tomography based on optical parametric amplification followed by direct detection. The method is immune to detection inefficiency and loss, and suitable for broadband, spatially and temporally multimode quantum states. To prove the principle, we experimentally reconstruct the Wigner function of squeezed vacuum occupying a single mode of a strongly multimode state. We obtain a squeezing of $-7.5\pm 0.4$ dB and a purity of $0.91^{+0.09}_{-0.08}$ despite more than $97\%$ loss caused mainly by filtering. Theoretically, we also consider the reconstruction of a squeezed single photon - a bright non-Gaussian state. Due to strong multimode parametric amplification, the method allows for the simultaneous tomography of multiple modes. This makes it a powerful tool for optical quantum information processing.
\end{abstract}


\maketitle

Quantum states of light promise to revolutionize nowadays technologies such as information processing~\cite{Zhong2020Dec,Larsen2019Oct}, metrology~\cite{LIGOScientificCollaborationandVirgoCollaboration2017Oct,Thekkadath2020Oct}, and sensing~\cite{Pirandola2018Dec}. In order to explore their non-classical features, quantum state tomography is employed \cite{Lvovsky2009Mar}. In particular, one can retrieve full information about a quantum state through reconstructing the Wigner quasi-probability distribution~\cite{Hillery1984Apr}. The inevitable challenge for the experimental reconstruction of the Wigner function is the fragility of the quantum states to losses, including detection inefficiency. Losses disturb quantum features like squeezing~\cite{Breitenbach1997May}, negativity of the Wigner function~\cite{Kenfack2004Aug}, and superpositions in the phase space~\cite{Gerry1998Jun,Sychev2017Jun}, which leads to wrong state reconstruction. This is the case with the most common method of tomography, based on the homodyne detection of optical quadratures~\cite{Smithey1993Mar}. Another drawback of homodyne tomography is the impossibility to address simultaneously different modes of multimode radiation, a property that gets increasingly important for optical quantum information~\cite{Ra2020Feb}.

As an alternative to homodyne detection, several groups reconstructed the Wigner function from the measurement of photon-number parity~\cite{Wallentowitz1996Jun,Banaszek1996Jun}. This method requires photon-number resolving detectors. In particular, the loss-tolerant tomography of a single-photon state was achieved by using time-multiplexed detection scheme~\cite{Laiho2010Dec}. However, such detectors impose a limitation on the brightness of the examined state. 

Here we propose and experimentally demonstrate another method of Wigner-function tomography using direct detection. It is based on the fact that after sufficiently strong phase-sensitive parametric amplification, the photon number scales as the squared quadrature at the input of the amplifier~\cite{Shaked2018Feb}, the choice of the quadrature being determined by the phase of the pump. Recently, this fact was used to retrieve the quadrature variances of squeezed vacuum~\cite{Shaked2018Feb, Frascella2019Sep,Takanashi2020Nov,Nehra2022Sep}. Here we show that parametric amplification enables the complete quantum state reconstruction, including the tomography of non-Gaussian quantum states, and different modes of a multimode state. As a proof of principle, we reconstruct the Wigner function of a single-mode squeezed vacuum state filtered from multimode radiation, and we show the method is loss- and noise-tolerant. We reconstruct a nearly pure state despite very low detection efficiency, additionally reduced because of filtering, without any correction for the detection loss and noise. The only losses contributing are those before the amplification, and they can be minimized. Our method can be applied to the tomography of faint and bright non-Gaussian states, as well as to the simultaneous tomography of states occupying different spatiotemporal modes of a multimode state. 

\begin{figure}[t]
\centering
\includegraphics[width=0.75\linewidth]{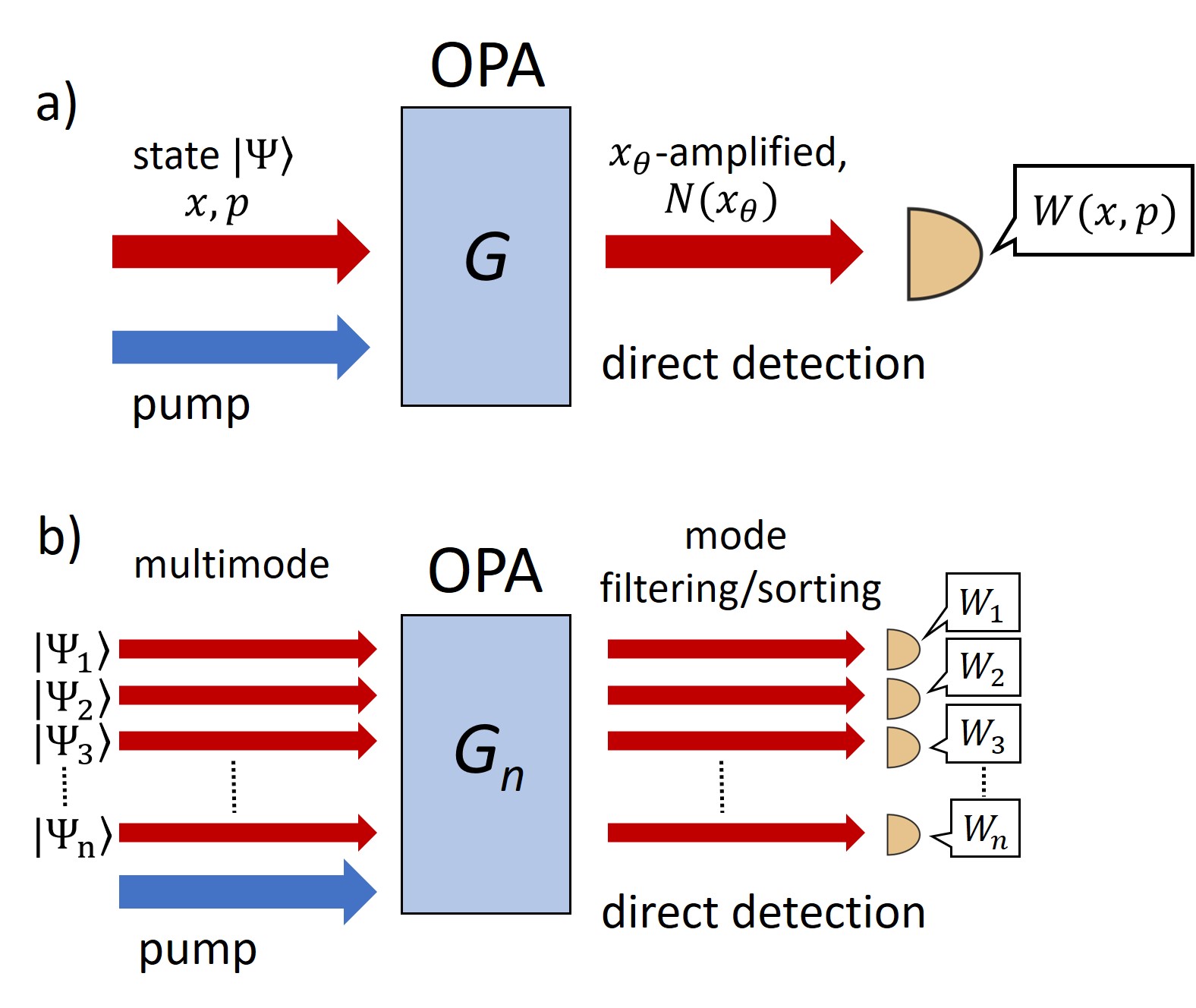}
     \caption{The idea of the method in single-mode (a) and multimode (b) cases. An optical parametric amplifier (OPA) amplifies a quadrature $x_\theta$ of the input quantum state and suppresses the conjugate quadrature. Under sufficient amplification gain $G$, only $x_\theta$ contributes to the photon number $N(x_\theta)$. Therefore, the statistics of $x_\theta$ can be retrieved via direct detection. The Wigner function $W(x,p)$ is then reconstructed from a set of $x_\theta$ distributions for different $\theta$.}
    \label{idea}
\end{figure}

Figure~\ref{idea}a shows the idea of our method. A quantum state $|\Psi\rangle$ is fed into a phase-sensitive optical parametric amplifier (OPA), with the squeezing parameter $G$. The OPA amplifies a certain quadrature $x_\theta=x\cos{\theta}+p\sin{\theta}$, where $x$ and $p$ are the position and momentum quadratures, and $\theta$ is the amplification phase being determined by the pump. The OPA amplifies $x_\theta$ by a factor $e^{G}$ and simultaneously de-amplifies the conjugate quadrature by the same amount. If $G$ is sufficiently high, the amplified quadrature $x_\theta$ dominates the output signal, and therefore, direct detection allows for extracting information about the quadrature $x_\theta$. 
Indeed, the photon-number operator after the amplification, $\hat{N}_\theta =\hat{x}'^2_\theta+\hat{p}'^2_\theta-\frac{1}{2}$, 
with $\hat{x}'_\theta=e^G \hat{x}_\theta$ and $\hat{p}'_\theta=e^{-G}\hat{p}_\theta$ being the output quadratures, under sufficiently high $G$ will have contribution only from the amplified quadrature,
\begin{equation}
    \hat{N}_{\theta} \approx
    e^{2G}\hat{x}_\theta^2 . \label{Homodyne N}
\end{equation}
At this point, the statistics of $N_{\theta}$ replicate those of $x_\theta^2$. We can thus obtain the continuous-variable 
probability distribution $P(|x_\theta|)$ (see Supplemental document 1B) and then, provided the input state is axially symmetric,  $P(x_\theta)=P(-x_\theta)$, also the complete quadrature probability distribution, 
\begin{equation}
    P(x_\theta)=e^G\sqrt{N_\theta} P(N_\theta).\label{Probability transformation}
\end{equation} 

With this approach, the losses present after amplification will not affect the obtained quadrature distributions, since the detection inefficiency just scales down the photon numbers, preserving the envelope of their distribution. Therefore, provided the quadratures are sufficiently amplified before being disturbed, quadrature distributions can still be retrieved no matter what optical losses and detection inefficiencies are present after amplification. 

 The detection scheme does not require photon-number resolution; after sufficient amplification, the photon number is measured as intensity, by photocurrent or charge integrating detectors like p-i-n diodes, charge-coupled devices (CCDs), or spectrometers. Such detection schemes smoothen the photon -number distribution~\cite{PhysRevA.88.023822}, turning the photon number into a continuous variable.
Typical values of the squeezing parameter in setups using strongly pumped parametric down-conversion (PDC) can reach $G=15$~\cite{Iskhakov2012Jun}. This value enables amplification by more than ten orders of magnitude, although in practice, even three to four orders suffice. For this reason, the detection noise is not a restriction either. It follows that the brightness of the state under study is not a limitation for this tomography scheme; it can be applied to both bright and faint states down to the single-photon level.

Importantly, the method can work even with broadband and multimode states, both spatially and temporally, since parametric amplification is intrinsically a multimode process~\cite{Wasilewski2006Jun,Sharapova2015Apr,LaVolpe2020Apr} (Fig.~\ref{idea}b). This can be achieved by tailoring the mode structure of the parametric amplifier to cover the mode content of the input states. If the input and amplifier modes match, the amplification can be simultaneous over all modes, and different quadratures $x_{\theta n}$ can be amplified depending on the phase between the pump and the modes. In this case, each mode will experience a certain amplification gain $G_n$. To retrieve the amplifier eigenmodes and the corresponding gain values, one only needs to amplify the  vacuum~\cite{Finger2017May,Averchenko2020Nov}. After the simultaneous multimode amplification, modes can be filtered out or sorted. 
Notably, this property is impossible with usual homodyne tomography.

\begin{figure}[h]
\includegraphics[width=0.95\linewidth]{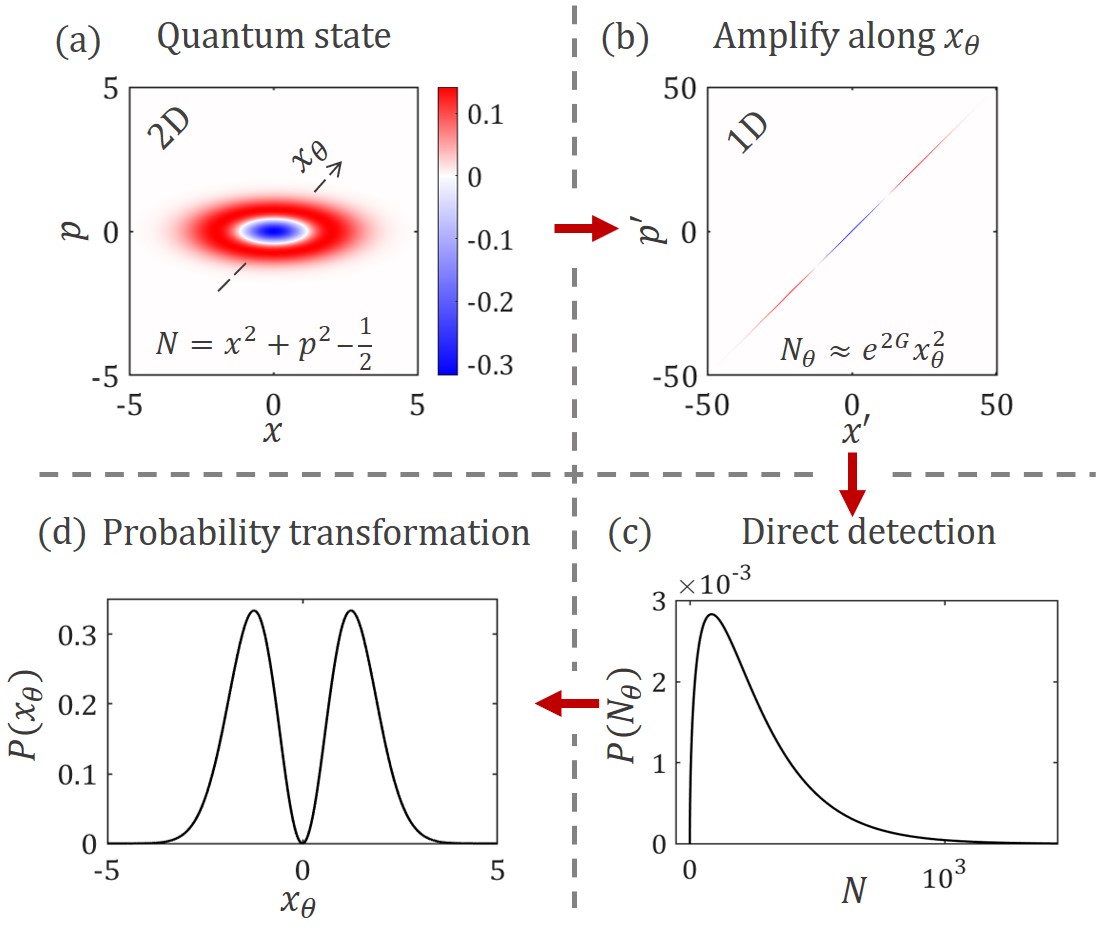}
     \caption{Calculated scenario of the Wigner-function tomography via parametric amplification: the case of a $4.3$ dB squeezed single photon. (a) The Wigner function of the input quantum state. (b) The Wigner function after the amplification of quadrature $x_\theta$, here $\theta=\pi/4$. Strong amplification ($G=2.7$) results in almost one-dimensional Wigner function, and the photon number is given by only the amplified quadrature.  (c) Continuous-variable photon-number distribution $P(N_\theta)$, to be measured via direct detection. (d) The quadrature distribution $P(x_\theta)$ retrieved via Eq.~(\ref{Probability transformation}).}
    \label{scheme}
\end{figure}

Calculations in Fig.~\ref{scheme} illustrate this procedure for the case of a non-Gaussian initial state: a $4.3$ dB squeezed single photon (panel a). After a sufficiently strong phase-sensitive amplification of quadrature $x_\theta$, the Wigner function $W(x,p)$ gets stretched along $x_\theta$. Figure~\ref{scheme}b shows the Wigner function after the amplification with the squeeze factor $G=2.7$ and a phase $\theta=\pi/4$. Despite a moderate, and definitely achievable in experiment, squeeze factor, the Wigner function becomes so stretched that the photon-number distribution is now fully determined by the one-dimensional marginal probability distribution of the amplified quadrature~\cite{Leuchs2015Jul}. 
In other words, the squared amplified quadrature can be mapped to the photon number,
\begin{equation}
    P(x'_\theta)=\int W(x'_\theta,p'_\theta)dp'_\theta,
    \label{marginal distribution}
\end{equation} 
and the probability distributions of the two variables are related through Eq.~(\ref{Probability transformation}). Figure~\ref{scheme}c shows the photon-number distribution of the amplified state, calculated using the continuous-variable approximation for the expression from Ref.~\cite{Kim1989Jul} (see Supplemental document 2). This distribution, measurable via direct detection, carries the information about marginal distribution (\ref{marginal distribution}), and therefore, the initial marginal distribution $P(x_\theta)$. The resulting marginal distribution, calculated using Eq.~(\ref{Probability transformation}), is shown in Fig.~\ref{scheme}d. Similar to the homodyne tomography of the Wigner function, a set of such marginal distributions for different phases $\theta$ enables the reconstruction of $W(x,p)$.

\begin{figure}[b]
\includegraphics[width=1\linewidth]{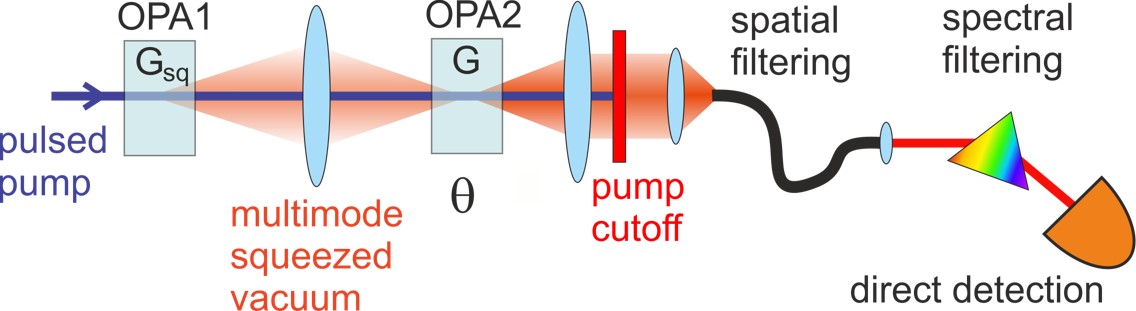}
\caption{Simplified experimental scheme: OPA1 generates broadband multimode squeezed vacuum, and for each mode, OPA2 amplifies some quadrature $x_\theta$, whose choice is determined by the pump phase. The amplified state is directly detected after spectral and spatial filtering for a set of $\theta$ values.}
\label{SV_idea}
\end{figure}

As a proof of principle, we reconstruct the Wigner function of a squeezed vacuum (SV) occupying one mode out of a highly multimode state, which is Gaussian but still very sensitive to losses. We use (Fig.~\ref{SV_idea}) two optical parametric amplifiers (OPAs) based on parametric down-conversion. OPA1 generates SV, with the squeezing parameter $G_{sq}$, and OPA2 performs phase-sensitive amplification with the squeezing parameter $G$. Both OPAs are pumped with picosecond pulses at $354.67$ nm, and both are highly multimode and broadband, unlike in the case of cavity or waveguide-based sources: the spectral bandwidth in the collinear direction is 30 nm and the angular divergence is 25 mrad. The SV emitted by OPA1 has 50 spatial and 370 spectral modes. To overcome the diffraction and make all spatial modes amplified, the SV emitted by OPA1 is imaged on OPA2~\cite{Frascella2019Sep}. 

The pump phase before OPA2, determining which quadrature $x_\theta$ is amplified, can be locked at different values (see Supplemental document 3A,E). To make Eq.~(\ref{Homodyne N}) valid for all phases $\theta$, the squeezed quadrature should be amplified sufficiently in order to overcome the initially anti-squeezed quadrature. Therefore, the choice of $G$ relies on the initial squeezing given by $G_{sq}$. In experiment, we set $G_{sq}=1.0\pm0.1$ and $G=4.4\pm0.1$, which makes Eq.~(\ref{Homodyne N}) valid to an accuracy better than 0.2\% (see Supplemental document 1A). In addition, this value of $G$ provides acceptable signal-to-noise ratio at the detection stage when amplifying the squeezed quadrature. 

After blocking the pump radiation with a dichroic mirror, we filter the SV spatially and spectrally. For spatial filtering, we couple the fundamental squeezed mode (whose shape is close to Gaussian) into a single-mode optical fiber (see Supplemental document 3B). The spectral filtering, with a monochromator, is to a bandwidth of $0.08$ nm at the degenerate wavelength $709.33$ nm. In addition to selecting just a fraction of the squeezing mode, it introduces more than 95\% loss (see Supplemental document 3C,D), which, however, does not affect the measurement due to the sufficiently strong amplification. Alternatively, multiple spatial and spectral modes can be sorted out and addressed simultaneously, by introducing a spatial light modulator and/or other optical elements~\cite{Defienne2020Jun,Reddy2014May,Zhou2017Dec,Gu2018Mar}.

Finally, the filtered radiation is detected with a triggered sCMOS camera (quantum efficiency 70\%). Out of the illuminated region, a single pixel is used, with dark counts of $2\pm1$ photons per pulse. This noise, although quite low, exceeds the mean photon number of the state, $\langle N\rangle=1.4$. But parametric amplification, similar to the local oscillator in homodyne detection, provides enough energy to overcome this noise. Overall optical losses after amplification exceed $97\%$ (see Supplemental document 3F).

We measure the number of photons pulse by pulse and acquire statistics over $8000$ pulses for different amplification phases ranging from $\theta=0$ (anti-squeezed quadrature $x$ amplified) to $\theta=\pi/2$ (squeezed quadrature $p$ amplified). In order to calibrate the measurement, we send to OPA2 the vacuum state, by simply blocking the SV radiation from OPA1.

\begin{figure}[t]
    \includegraphics[width=1\linewidth]{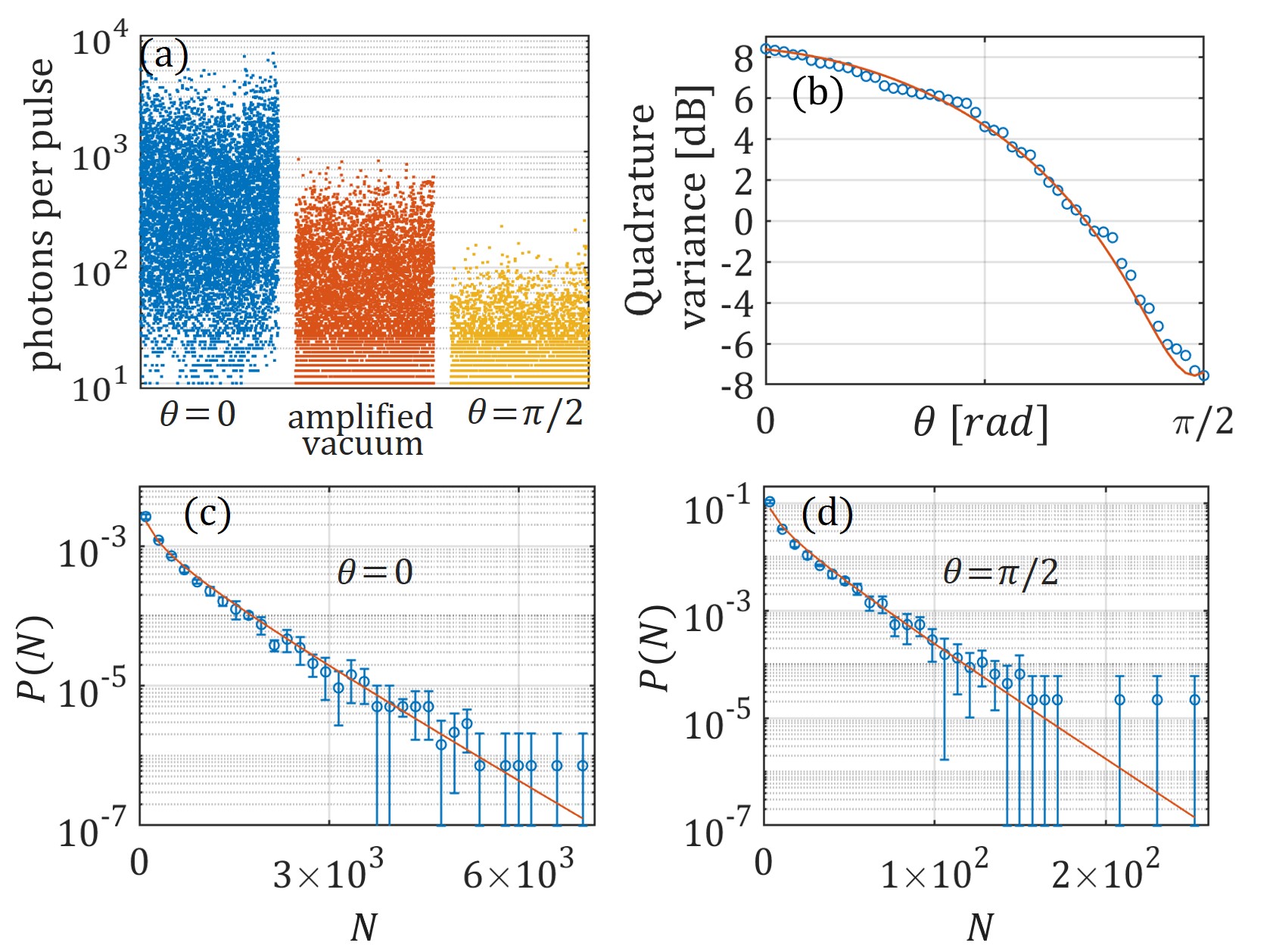}
     \caption{(a) Sets of 8000 photon-number measurements for the cases of amplifying the anti-squeezed quadrature (blue), the vacuum (orange) and the squeezed quadrature (yellow). (b) Measured quadrature variance as a function of $\theta$ (points) and its sinusoidal fit (line). (c), (d) Photon-number distributions for the cases of amplifying anti-squeezed and squeezed quadratures, respectively (points) and their fits (lines).}
\label{measurements}
\end{figure}

Figure~\ref{measurements}a shows the results of these photon-number measurements for different experimental settings. In one measurement, OPA2 amplified the anti-squeezed quadrature (blue points). In another case, OPA2 amplified the vacuum because SV after OPA1 was blocked (orange points). In the third case, OPA2 amplified the squeezed quadrature (yellow points). The measured mean photon numbers in these three cases are $\langle\hat{N}_0\rangle=511\pm7$, $\langle\hat{N}_{vac}\rangle=73\pm1$, and $\langle\hat{N}_{\frac{\pi}{2}}\rangle=12.8\pm0.2$ photons, respectively. 

Recalling that the mean photon numbers after the amplification scale as the squared quadratures before the amplification, see Eq.~(\ref{Homodyne N}), and for the SV state, $\langle\hat{x}_\theta\rangle=0$, the mean photon number after OPA2 is a measure of the quadrature variance at its input~\cite{Shaked2018Feb}. 

The quadrature variance, normalized to the vacuum level, is shown in Fig.~\ref{measurements}b (points) as a function of $\theta$, fitted by $Var(\hat{x}_\theta)/Var(\hat{x}_{vac})=a\cos^2{\theta}+d$ (line). From this dependence, we obtain the degrees of squeezing $-7.5\pm0.2$ dB and anti-squeezing $8.4\pm0.1$ dB. The measured squeezing is in good agreement with the values of $G_{sq}$, optical losses ($0.6\%$) before amplification and the imperfect alignment/mode matching (characterized by the visibility of phase-sensitive amplification $V=95\%$), which adds $5.3\%$ to the total amount of loss (see Supplemental document 1D,3F). 
\begin{figure}[b]
    \includegraphics[width=1\linewidth]{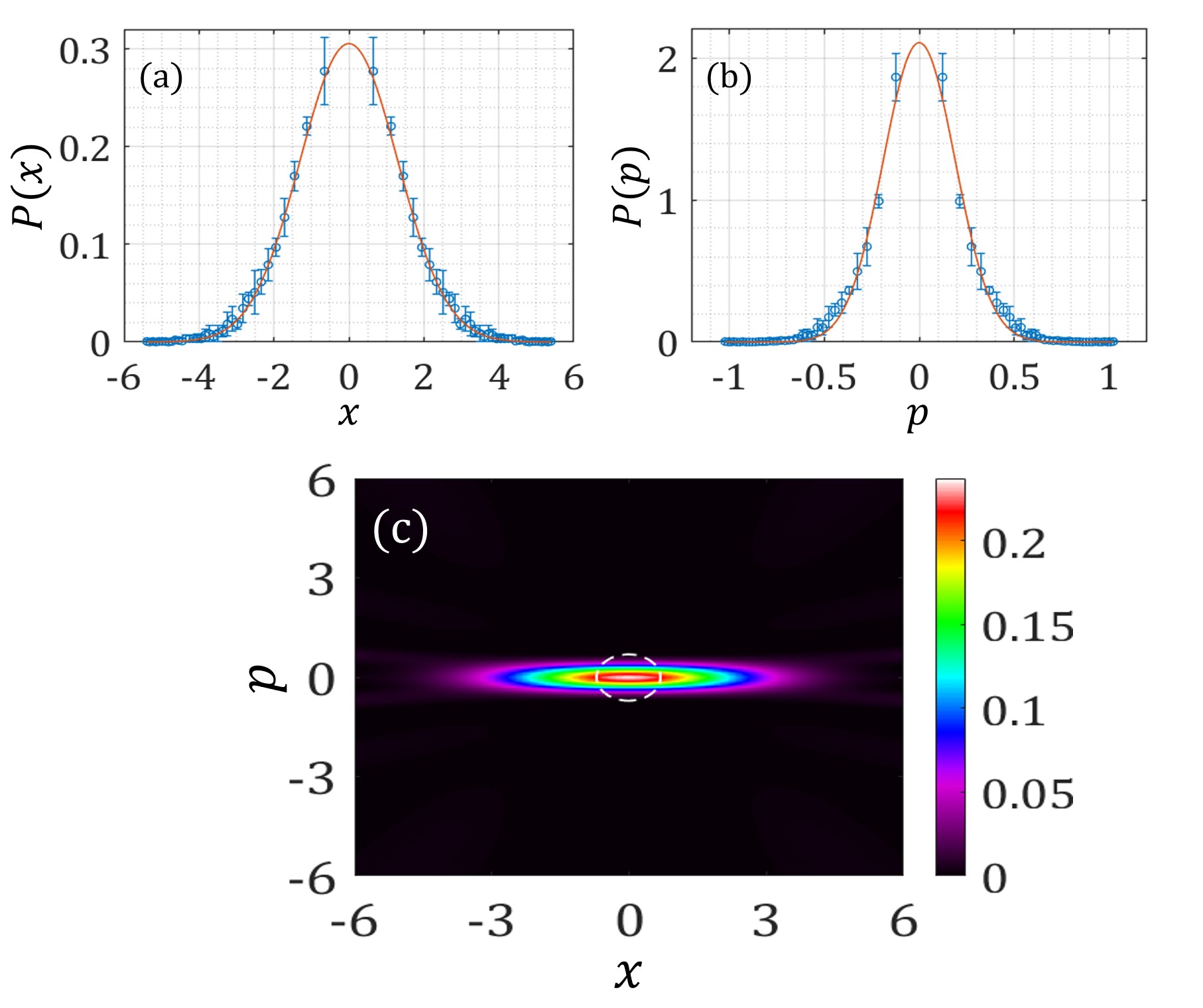}
     \caption{\label{results}(a), (b) Measured quadrature distributions for the anti-squeezed and squeezed quadratures, respectively (points) and their Gaussian fits (lines). (c) Reconstructed Wigner function of the SV state. White dashed line shows the Wigner function of the vacuum state at $1/\sqrt{e}$ level. }
\end{figure}

We obtained photon-number probability distributions by sampling the photon-number data into $35$ bins for each phase value.
Figures~\ref{measurements}c,d show these distributions for the cases of amplified anti-squeezed and squeezed quadratures, respectively (points). Due to imperfect spatial filtering, the detected number of modes was $1.2$, which was taken into account in the fit (lines); see Supplemental document 1C,3D.

The quadrature probability distributions were obtained by applying transformation~(\ref{Probability transformation}) to the corresponding measured photon-number distributions. Figures~\ref{results}a,b show the examples for the anti-squeezed and squeezed quadratures, respectively. 

Finally, we reconstructed the Wigner function of the squeezed vacuum state by applying the inverse Radon transform to the obtained fits of all quadrature probability distributions \cite{Lvovsky2009Mar}. The reconstructed distribution (Fig.~\ref{results}c) fairly resembles a squeezed vacuum state with $\Delta x=1.30\pm0.06$ and $\Delta p=0.21\pm0.01$. The dashed white circle at the center marks the Wigner function of the vacuum state at $1/\sqrt{e}$ level, corresponding to $\Delta x_{vac}=0.5$. The reconstruction yields the amounts of squeezing and anti-squeezing of $-7.5\pm0.4$ dB and $8.3\pm0.4$ dB, respectively, in perfect agreement with the values obtained by measuring the mean photon number. The purity of the state~\cite{Adesso2014Mar} was found to be $\Delta^2 x_{vac}/(\Delta x \Delta p)=0.91^{+0.09}_{-0.08}$. The fidelity of this state to the SV state calculated theoretically for $G_{sq}=1$ is $99.4\%$. These results, obtained without any correction for losses, prove the feasibility of the method under real-life conditions.

In conclusion, we have demonstrated the tomography of quantum states based on optical parametric amplification, which provides its tolerance to detection loss and noise. As a proof of principle, we applied the method to a squeezed vacuum state, achieving a degree of squeezing $-7.5\pm0.4$ dB and a purity of $0.91^{+0.09}_{-0.08}$, despite more than 97\% losses in the detection channel. With such losses, almost no squeezing could be observed with homodyne detection. 

The method can also be applied to non-Gaussian states, including bright ones, which are especially susceptible to losses. The only restriction is that the state should have axially symmetric Wigner function, but this includes a vast variety of non-Gaussian states, in particular, Fock, squeezed Fock, even/odd Schr\"odinger cat, and importantly, the GKP states~\cite{Gottesman2001Jun} which are required for fault-tolerant quantum computing~\cite{Baragiola2019Nov}. 

 Parametric amplification is a multimode process, therefore, the method is suitable for the tomography of broadband and multimode quantum states. A set of multiple input modes can be amplified simultaneously if they match the amplifier eigenmodes. Afterwards, a specific mode can be filtered out. In our experiment we filtered a single spatial eigenmode from a set of 50 modes, and a fraction of a frequency eigenmode out of 370 modes.   
 
More interestingly, being immune to loss, the method also allows for the simultaneous tomography of multiple modes, if a mode sorter is placed before detection. By scanning the phase between the pump and the modes, different quadratures of all modes can be measured simultaneously. Notably, this property is impossible with usual homodyne tomography because of inevitable losses accompanying mode sorting.

The possibility to characterize all modes at once and the loss immunity makes the method a perfect candidate for high-dimensional quantum information applications~\cite{Ra2020Feb,Erhard2020Jul}. Moreover, the method can assist chip-based generation, manipulation, and detection of quantum states~\cite{Nehra2022Sep,Yang2021Aug,Zhang2019Dec,Masada2015May,Raffaelli2018Feb}, which paves the way towards real photonic quantum computers. 

\textbf{Acknowledgments} This work was funded within the QuantERA II Programme (project SPARQL) that has received funding from the European Union’s Horizon 2020 research and innovation programme under Grant Agreement No 101017733, with the funding organization Deutsche Forschungsgemeinschaft. 

\textbf{The authors} are part of the Max Planck School of Photonics, supported by BMBF, Max Planck Society, and Fraunhofer Society. We thank Farid Khalili, Radim Filip, Laszlo Ruppert and \'Eva R\'acz for helpful discussions.

\textbf{Disclosure} The authors declare no conflicts of interest.

\textbf{Supplemental document}
See Supplemental document for the derivation of Eq. \ref{Probability transformation}, more experimental details, and the relation between the visibility and efficiency of the proposed scheme.

\bibliography{References}


\end{document}